\documentclass[a4paper,11pt]{article}
\usepackage{pos}

\title{Classical vs. quantum corrections to jet broadening in a weakly coupled QGP}
\author*[a]{Eamonn Weitz}

\affiliation[a]{SUBATECH, Nantes Universit\'e, IMT Atlantique, IN2P3/CNRS,\\
4 rue Alfred Kastler, La Chantrerie BP 20722, 44307 Nantes, France\\}

\emailAdd{eamonn.weitz@subatech.in2p3.fr}

\abstract{
We compute double-logarithmically enhanced corrections to $\hat{q}$ at relative order $\mathcal{O}(g^2)$ in the setting 
of a weakly coupled quark-gluon plasma, observing how the thermal scale affects the region of phase space, which gives rise 
to these corrections. 
We furthermore clarify how the region of phase from which these 
corrections are borne is situated with respect to that from which the 
classical corrections arise at relative order $\mathcal{O}(g)$.
This represents a significant step towards our eventual goal of understanding which class
of corrections dominate, thereby pushing forward
our quantitative grasp on the phenomenon of jet quenching in heavy-ion collisions.}

\FullConference{HardProbes2023\\
 26-31 March 2023\\
 Aschaffenburg, Germany\\}

\begin{document}
\maketitle
\section{Introduction}
In the context of heavy-ion collisions, jets provide an ideal \emph{hard probe} of the quark-gluon plasma (QGP). Through interacting 
with the QGP, they receive momentum kicks in the directions transverse to their propagation -- \emph{transverse momentum broadening}.
This broadening can be captured by the \emph{transverse momentum broadening 
coefficient}, $\hat{q}=\langle k_{\perp}^2\rangle /L$, which 
specifies the transverse momentum picked up per unit length, $L$ by a hard parton propagating through the QGP. See 
the recent reviews on jets in heavy-ion collisions \cite{Apolinario:2022vzg} or 
extractions of $\hat{q}$ from data \cite{Han:2022zxn} for more information.

 For a weakly coupled QGP, $\hat{q}$ can be expressed in terms of the 
 transverse scattering kernel
 \begin{equation}
    \hat{q}(\mu)=\int^{\mu}\frac{d^2 k_{\perp}}{(2\pi)^2}k_{\perp}^2\mathcal{C}(k_\perp),
    \quad\quad \mathcal{C}(k_\perp)\equiv(2\pi)^2\frac{d\Gamma}{dk_{\perp}^2},
 \end{equation}
 where $\frac{d\Gamma}{dk_{\perp}^2}$ is the rate for a hard parton with energy, $E\gg T$, the temperature of the plasma, propagating 
 along the $z$ direction to pick up $k_{\perp}$. The cutoff, $\mu$ 
 is installed so as not to include larger momentum scatterings, which include two 
 hard partons in the final state. See App. A of \cite{Ghiglieri:2022gyv} for our conventions.
 
 At leading order (LO) in $g$, $\hat{q}$ receives contributions from the hard $(T)$ \cite{Arnold:2008vd} and soft $(gT)$ \cite{Aurenche:2002pd} scales, which 
 give rise to the parametric form (up to logarithms) $\hat{q}\sim g^4 T^3$. The soft contribution 
 is cut off in the IR by dynamical screening, implemented through 
 Hard Thermal Loop Effective Theory (HTL) resummation \cite{Braaten:1991gm}. NLO corrections also come from the soft scale \cite{CaronHuot:2008ni}.
Ultrasoft $(g^2T)$ modes contribute at $\mathcal{O}(g^2)$, for which the perturbative expansion breaks down. We 
refer to these NLO and NNLO contributions as \emph{classical corrections}: they are distributed on the $T/\omega$ IR tail 
of the Bose-Einstein distribution, $n_{\mathrm{B}}(\omega)$ and are therefore sourced by the Matsubara zero-mode.

Caron-Huot \cite{CaronHuot:2008ni} demonstrated that one may compute the zero-mode contribution 
to $\mathcal{C}(k_{\perp})$ in Electrostatic QCD (EQCD) \cite{Braaten:1995cm}, meaning that one 
can bypass the somewhat cumbersome HTL computation. More importantly, 
as a theory of static modes, EQCD is amenable to study using three-dimensional lattice simulations, which can thus 
provide a \emph{non-perturbative} evaluation of $\mathcal{C}(k_{\perp})$, summing contributions 
from the soft and ultrasoft scales to all orders \cite{Panero:2013pla,Moore:2019lgw}. Recently, the impact 
of these classical corrections on the in-medium splitting rate was assessed \cite{Moore:2021jwe,Schlichting:2021idr} and 
found to be very relevant. A 
similar program is well underway for the non-perturbative determination 
of classical corrections to the \emph{asymptotic mass} \cite{Moore:2020wvy,Ghiglieri:2021bom,Ghiglieri:2023cyw}.

These classical corrections are at odds with doubly-logarithmically enhanced radiative, \emph{quantum corrections}, 
appearing at $\mathcal{O}(g^2)$, first identified in
\cite{Liou:2013qya,Blaizot:2013vha}. There, the leading enhancement is
 $\sim\ln^2 L_{\text{med}}/\tau_{\text{min}}$, with $L_{\text{med}}$ the length 
of the medium and $\tau_{\text{min}}\sim 1/T$ the 
minimum formation time of the associated radiation. This potentially 
large double-logarithm can be resummed \cite{Blaizot:2014bha}, with the evolution equations 
solved numerically in \cite{Caucal:2021lgf,Caucal:2022fhc}. Interestingly, they 
also arise in the context of double gluon emission \cite{Blaizot:2014bha,Arnold:2015qya}, implying 
that these logarithms are subject to a certain universality.

These corrections come from the 
\emph{single scattering regime} where bremsstrahlung is sourced by a single scattering 
with the medium. This is in contrast to the \emph{multiple scattering regime}, where the bremsstrahlung's 
formation time, $\tau$ is long enough so that it is coherently triggered by multiple collisions, accounted for through
 LPM resummation. In \cite{Ghiglieri:2022gyv}, we compute 
these doubly-logarithmically enhanced corrections in the context of a weakly coupled QGP, carefully analysing how 
the thermal scale deforms the region of phase space from which the double-logarithms emerge.
\begin{figure}[t]
    \centering
      \includegraphics[width=0.7\textwidth]{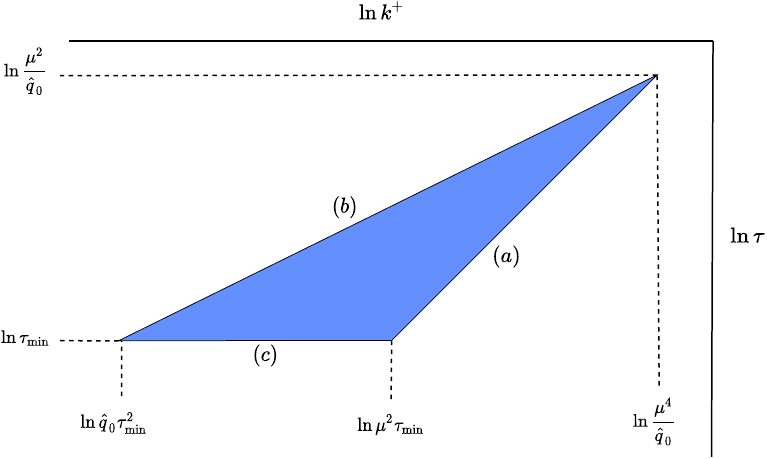}%
    \caption{Depiction of bounds from the integration in Eq.~\eqref{eq:dlog_lit}. The $(b)$ boundary is defined by $\tau=\sqrt{k^+/\hat{q}_0}$ and 
    the $(a)$ boundary by $\tau=k^+/\mu^2$. Figure taken from \cite{Ghiglieri:2022gyv}.} 
    \label{fig:bdimtriangle}
    \end{figure}

\section{Double Logarithmic Corrections and the Thermal Scale}
The correction from \cite{Liou:2013qya} emerges in the standard dipole picture 
\begin{equation}
    \delta\hat{q}_{\text{\cite{Liou:2013qya,Blaizot:2013vha}}}(\mu)=4\alpha_sC_{R}\hat{q}_0\int^{\mu}\frac{d^2k_{\perp}}{k_{\perp}^2}\int\frac{dk^+}{k^+},
\end{equation}
where $k^+\equiv k_{\perp}^2\tau$ is the energy of the bremsstrahlung and $\hat{q}_0$ is the LO transverse momentum broadening coefficient, stripped 
of the Coulomb logarithm as is done in the \emph{harmonic oscillator approximation} (HOA). 
Here we can explicitly see that one of the logarithms comes from a soft, $dk^+/k^+$ divergence, 
with the other 
coming from a collinear, $d^2 k_{\perp}/k_{\perp}^2$ divergence. For what follows, it turns out to be more 
convenient to work with $\tau$ and $k^+$
\begin{equation}
    \delta\hat{q}_{\text{\cite{Liou:2013qya,Blaizot:2013vha}}}(\mu)=\frac{\alpha_s C_R}{\pi}\hat{q}_0\int_{\tau_{\text{min}}}^{\mu^2/\hat{q}_0}\frac{d\tau}{\tau}
    \int_{\hat{q}_0\tau^2}^{\mu^2\tau}\frac{dk^+}{k^+}
    =\frac{\alpha_s C_R}{2\pi}\hat{q}_0\ln^2\frac{\mu^2}{\hat{q}_0\tau_{\text{min}}}.
    \label{eq:dlog_lit}
\end{equation}
 The limits above
come from integrating over the triangle presented in Fig.~\ref{fig:bdimtriangle}. Boundary $(a)$ arises from 
the need to cut off non-diffusive momentum exchanges above the scale $\mu$. The line $(b)$ is then 
defined by $k_{\perp}^2\equiv\hat{q}_0\tau$, marking the boundary with the \emph{deep LPM regime} in which multiple scatterings 
occur. Above boundary $(b)$ there is no longer a double-logarithmic enhancement 
as the $k^+$ integrand changes as $1/k^+\rightarrow 1/\sqrt{k^+}$. Finally, boundary $(c)$ is 
an artefact of the \emph{instantaneous approximation}: scatterings between the jet and medium are assumed 
to take place instantaneously compared to the formation time associated with the radiation. The 
result from \cite{Liou:2013qya} is recovered upon identifying $\mu$ with 
the \emph{saturation scale}, $Q_s\equiv\hat{q} L_{\text{med}}$.
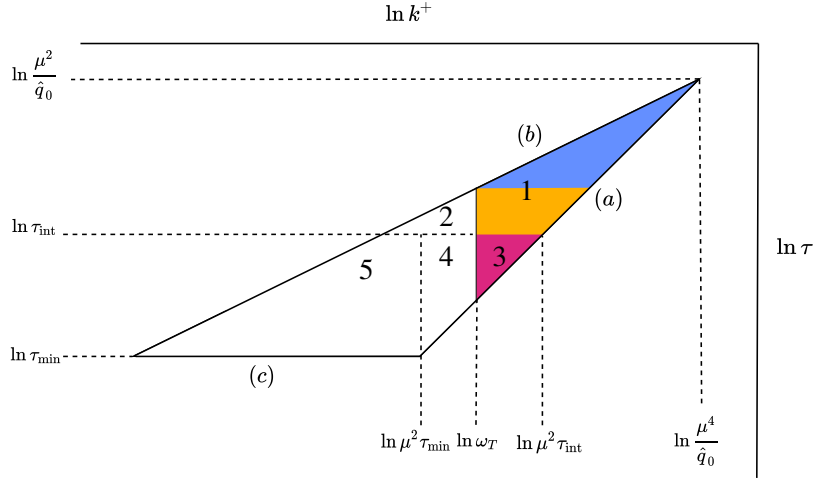
\begin{figure}[t]
    \centering
      \includegraphics[width=0.7\textwidth]{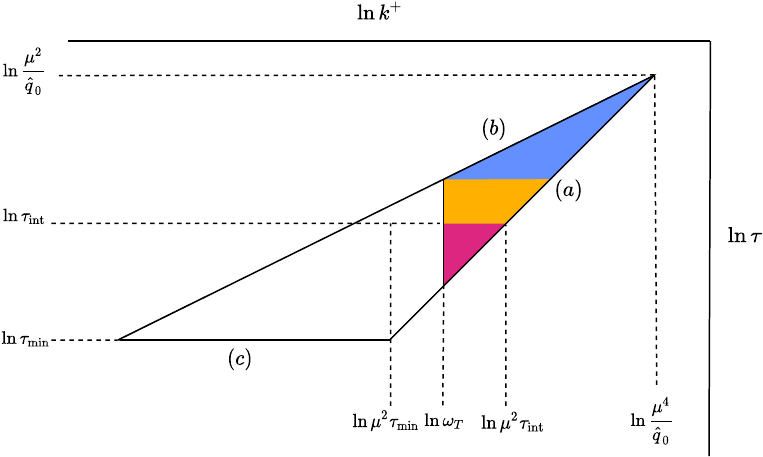}
      \put(-110,105){1}
      \put(-140,95){2}
      \put(-120,80){3}
      \put(-140,80){4}
      \put(-170,75){5}
    \caption{Deformation of the double-logarithmic phase space with the inclusion of thermal effects. Regions 
    1 and 2 form the ``few scattering'' regime, over which we integrate to get Eq.~\eqref{eq:BDIMnonzeroTred} whereas
    we integrate over regions 3 and 4, the ``strict single scattering'' regime to get Eq.~\eqref{eq:sss_before}. Region 
    5 then gives rise to the $\mathcal{O}(g)$ corrections to $\hat{q}$, calculated in \cite{CaronHuot:2008ni}. Figure taken 
    from \cite{Ghiglieri:2022gyv}.} 
    \label{fig:triangle}
\end{figure}

In a weakly coupled QGP, as soon as the energy overlaps with the temperature scale, one needs to account for 
more medium effects than those captured by these instantaneous, spacelike interactions. Specifically, by taking $T>\mu\gg\sqrt{g} T$\footnote{The 
$\mu>T$ case is studied in \cite{Ghiglieri:2022gyv}.} and
replacing $1\rightarrow(1+2n_{\text{B}}(k^+))$ in the $k^+$ integrand of Eq.~\eqref{eq:dlog_lit}, we find
\begin{equation}
        \delta \hat{q}(\mu)^{\mathrm{few}}=\frac{\alpha_s C_R}{2\pi}\hat{q}_0\bigg\{\ln^2\frac{\mu ^2}{\hat{q}_0 \tau_{\mathrm{int}}} 
        -\frac{1}{2}\ln^2\frac{\omega_{\mathrm{T}}}{\hat{q}_0\tau_{\mathrm{int}}^2}
        \bigg\}\quad \text{with}\;\omega_{\text{T}}=\frac{2\pi T}{e^{\gamma_E}}\quad \text{for}\;\;  
        \frac{\omega_{\mathrm{T}}}{\mu^2}\ll\tau_{\mathrm{int}}\ll\sqrt{\frac{\omega_{\mathrm{T}}}{\hat{q}_0}}.
        \label{eq:BDIMnonzeroTred}
\end{equation}
In doing so, we account for the Bose-Einstein stimulated emission of the radiated gluon as 
well as the absorption of a gluon from the medium. We will comment shortly on the purpose of $\tau_{\text{int}}$. 
But can these additional effects be neglected in a way that is 
consistent with single scattering, for instance, by demanding that $\hat{q}_0\tau^2_{\text{min}}\gg T$ in Eq.~\eqref{eq:dlog_lit}\footnote{This demand 
is motivated by the fact that $n_{\text{B}}(k^+)$ is exponentially suppressed for $k^+\gg T$.}? 
It turns 
out that the answer is no \cite{Ghiglieri:2022gyv}: such a choice of $\tau_{\text{min}}$ would necessarily allow for formation times 
associated with the deep LPM regime, where $\tau_{\mathrm{LPM}}\gtrsim 1/g^2T$. 
Note that the correction in Eq.~\eqref{eq:BDIMnonzeroTred}
corresponds to integrating over the 1 and 2 regions in Fig.~\ref{fig:triangle}.

The requirement $1/g^2T\gg\tau_{\mathrm{int}}\gg 1/gT$
means that processes where a \emph{few scatterings} occur are included in Eq.~\eqref{eq:BDIMnonzeroTred}. Indeed, $\tau_{\mathrm{int}}$
defines a border with what we have identified as a \emph{strict single scattering} regime, where the formation time is \emph{a priori}
consistent with single scattering, i.e $\tau\ll 1/g^2 T$. This region is characterised by so-called 
\emph{semi-collinear} processes \cite{Ghiglieri:2015ala}, where timelike as 
well as spacelike exchanges are allowed to occur. The leading contribution from this region is 
given by integrating over the 3 and 4 regions in Fig.~\ref{fig:triangle} and yields
\begin{equation}
    \delta\hat{q}_{\text{semi}}(\mu)=\frac{\alpha_s C_{R}}{2\pi}\hat{q}_0\ln^2\frac{\mu^2\tau_{\text{int}}}{\omega_{\mathrm{T}}},
    \label{eq:sss_before}
\end{equation}
where we have taken the HOA. Adding Eqs.~\eqref{eq:BDIMnonzeroTred}, ~\eqref{eq:sss_before}, we then find 
\begin{equation}
    \delta\hat{q}(\mu_{\perp})_{\text{dlog}}=\frac{\alpha_s C_{R}}{4\pi}\hat{q}_0\ln^2\frac{\mu^4}{\hat{q}_0\omega_\mathrm{T}}.\label{eq:dlog_final}
\end{equation}
 As well as the disappearance of $\tau_{\mathrm{int}}$, we note the absence of an IR cutoff, $\tau_{\mathrm{min}}$; looking 
 to Fig.~\ref{fig:triangle}, the double-logarithm is instead cut off by the scale $\omega_{\mathrm{T}}$. Thus, the 
 thermal scale plays an extremely important role in this context.

 \section{Relation to Classical Corrections}
As well as double-logarithmic corrections at $\mathcal{O}(g^2)$, we also find \emph{power law 
corrections} when integrating over regions 3 and 4
\begin{equation}
    \delta{q}_{\mathrm{PL}}=\frac{\alpha_s C_R }{2\pi} \hat{q}_0
    \frac{4T \ln \left(\frac{\mu^2
   \tau_{\text{int}}}{k^{+}_{\text{IR}} e}\right)}{k^{+}_{\text{IR}}},
   \label{eq:semievaltaufinal}
\end{equation}
where $k^{+}_{\mathrm{IR}}$ is an IR cutoff on the energy. Power law corrections 
of this kind are usually discarded as they are unphysical --  they always cancel against 
other power law corrections coming from adjacent regions of phase space. Here, we use this 
fact to our advantage; in the calculation of the $\mathcal{O}(g)$ corrections, power law corrections should 
appear, with $k^{+}_{\mathrm{IR}}$ instead acting as a UV cutoff there. 

In more detail, one can use causality properties of $\mathcal{C}(k_{\perp})$, also revealed in \cite{CaronHuot:2008ni}, 
to carry out the $k^+$ integral by analytically continuing into the $k^+$ complex plane. $k^{+}_{\mathrm{IR}}$
then appears as the radius of the arc of the deformed contour, with the arc lying between the zeroth and 
first Matsubara modes. There is no dependence on $k^{+}_{\mathrm{IR}}$ in \cite{CaronHuot:2008ni} as 
the $1/k^{+}_{\mathrm{IR}}$ terms go to zero and can thus be safely neglected. Nevertheless, we have indeed 
computed these arc contributions explicitly and shown that they cancel exactly 
against the result from Eq.~\eqref{eq:semievaltaufinal}, further 
confirming how the region from which the classical corrections emerge 
is connected to that associated with the logarithmically-enhanced quantum corrections.

\section{Conclusion and Outlook}
We have studied how, in the setting of a weakly coupled QGP, the thermal scale 
affects the double-logarithmic phase space, originally identified in \cite{Liou:2013qya,Blaizot:2013vha}. In 
more detail, we showed how the scale, $\omega_{\mathrm{T}}$ cuts off this region of phase space
and furthermore, how the region, which gives rise to the classical corrections, computed in 
\cite{CaronHuot:2008ni} fits in comparison.

In obtaining Eq.~\eqref{eq:dlog_final}, we have taken the HOA, neglecting a neighbouring region 
phase space, which permits both single and multiple scattering processes. To 
properly deal with such a region, we would need to solve an LPM resummation equation, 
derived in \cite{Ghiglieri:2022gyv} (see also \cite{Iancu:2014kga}),
 differential in the transverse momentum picked up by the parton. We foresee that
the use of the \emph{improved opacity expansion} \cite{Barata:2021wuf} could allow us to 
arrive at an approximate solution of this equation but we leave such an endeavour to future work.

\section*{Acknowledgements}
We thank Jacopo Ghiglieri for collaboration on the original work 
\cite{Ghiglieri:2022gyv}.

\bibliographystyle{JHEP}
\bibliography{eloss.bib}

\end{document}